\documentclass[11pt]{article}

\usepackage[T2A,T1]{fontenc}
\usepackage[centertags]{amsmath}
\usepackage{amsfonts}
\usepackage{amssymb}
\usepackage[numbers,sort&compress]{natbib}
\usepackage[pdftex]{hyperref}
\usepackage[pdftex]{graphicx}
\usepackage[russian,english]{babel}

\usepackage{paperinitial}

\paperinitialization{15mm}{15mm}{20mm}{10mm}{2pt}{10pt}

\DeclareMathOperator{\sgn}{sgn}

\newcommand{\e}{\varepsilon}
\newcommand{\vf}{\varphi}

\newcommand{\al}{\alpha}

\newcommand{\ga}{\gamma}

\newcommand{\de}{\delta}

\newcommand{\la}{\lambda}

\newcommand{\ups}{\upsilon}

\begin{document}

\allowdisplaybreaks[4]
\selectlanguage{english}


\title{{\Large \textbf{Properties of electrons scattered on a strong plane electromagnetic wave with a linear polarization: classical treatment}}}

\date{}

\author{O.~V. Bogdanov\thanks{E-mail: \texttt{bov@tpu.ru}}\;  and P.~O. Kazinski\thanks{E-mail: \texttt{kpo@phys.tsu.ru}}\\[0.5em]
{\normalsize ${}^*$Department of Higher Mathematics and Mathematical Physics,}\\
{\normalsize Tomsk Polytechnic University, Tomsk, 634050, Russia}\\[0.3em]
{\normalsize ${}^\dag$Physics Faculty, Tomsk State University, Tomsk, 634050, Russia}
}

\maketitle

\begin{abstract}

The relations among the components of the exit momenta of ultrarelativistic electrons scattered on a strong electromagnetic wave of a low (optical) frequency and linear polarization are established using the exact solutions to the equations of motion with radiation reaction included (the Landau-Lifshitz equation). It is found that the momentum components of the electrons traversed the electromagnetic wave depend weakly on the initial values of the momenta. These electrons are mostly scattered at the small angles to the direction of propagation of the electromagnetic wave. The maximum Lorentz factor of the electrons crossed the electromagnetic wave is proportional to the work done by the electromagnetic field and is independent of the initial momenta. The momentum component parallel to the electric field strength vector of the electromagnetic wave is determined only by the diameter of the laser beam measured in the units of the classical electron radius. As for the reflected electrons, they for the most part lose the energy, but remain relativistic. There is a reflection law for these electrons that relates the incident and the reflection angles and is independent of any parameters.

\end{abstract}

\section{Introduction}

The scattering of electrons on the electromagnetic wave is a textbook subject both at classical and quantum level. Even in the regime of strong strengths, the transition amplitude in a tree approximation is well-known and calculated using the exact solutions to the Dirac equation (see, e.g., \cite{LandLifshQED}). However, for the laser radiation with large intensity (it should be of the order of $I\gtrsim10^{24}$ W/cm${}^2$ for the optical laser), the process of radiation of soft photons becomes relevant and even dominates at certain circumstances. Hence, an infinite number of diagrams have to be summed in order to describe correctly the properties of the scattered electrons and photons in this regime. Furthermore, at such intensities, the scattering amplitudes depend nonlinearly on the field strength of the electromagnetic wave \cite{Ritus} and so the results strongly depend on the shape of the laser beam. The lasers with such intensities will become accessible in the nearest future \cite{ELI,ECELS}, but a thorough theory of interaction of strong electromagnetic waves with a matter is far from being complete. We report on new (to our knowledge) startling features of the unpolarized electrons scattered on a laser beam with linear polarization. These properties, in particular, allow one to manipulate the characteristics of electron bunches, in addition to the standard means (see, e.g., \cite{HSXZ}).

The problem of direct summing an infinite number of diagrams with soft photons and dressed electron propagators is a formidable task. Therefore, we shall study this problem in the quasi-classical approximation neglecting the radiation of hard photons with the energies comparable to the electron energy, the vacuum polarization effects, and the electron-positron pairs production processes. This is a reasonable approximation for the electromagnetic field strengths less than the critical (Schwinger) field (the intensity $I\approx10^{30}$ W/cm${}^2$). Besides, we completely neglect the interaction of electrons in the bunch and suppose that the sizes of electron's wave packet in the laboratory frame are less than one hundredth of the laser wavelength during all the interaction process. Then it can be shown (see, e.g., \cite{BBT} and also \cite{JoHu}) that the center of an electron wave packet moves along the trajectory obeying the classical equations of motion (the Lorentz equation) with the electromagnetic field representing a superposition of the external field and the field created by an electron wave packet. If the wave packet is localized enough (as we have assumed), the standard arguments used in deriving the Lorentz-Dirac (LD) equation (see, e.g., \cite{Lor,Dir,Barut,RohrlBook,Spohn,Kos,siss}) transform this Lorentz equation into the LD equation in the leading nontrivial order in the size of the wave packet. So, in fact, we have to investigate the solutions to the LD equation. In the standard $S$ matrix approach, we need to sum an infinite number of diagrams with soft photons in order to describe such dynamics of the electron wave packet.

The quantum corrections due to radiation of hard photons can be approximately taken into account introducing an additional stochastic force to the equations of motion of the wave packet center \cite{SokTer}. This results in a broadening of the wave packet, but the dynamics of its center are still described well by the classical equations of motion under the assumptions made. Moreover, the late time asymptotics, we shall use in analyzing the properties of the scattered electrons, are attractors for the set of the physical solutions to the LD equation \cite{ldesol,rde}. Having emitted a hard photon and suffered a recoil, the electron tends to return to the initial trajectory (the attractor). The correction to the LD equation due to spin can be also neglected for the unpolarized wave packets of electrons with a vanishing average spin. The effect of radiative polarization is negligibly small in the ultrarelativistic case we consider (see \cite{TeBaKha} and for a recent discussion, e.g., \cite{KarlPRA}).

So, if we apply the LD equation to describe the dynamics of electrons in the field of a plane linearly polarized electromagnetic wave with the intensity $I\gtrsim10^{24}$ W/cm${}^2$ and the energy of photons $\Omega\approx1$ eV, we shall see that, in the ultrarelativistic limit, the electrons scattered on this wave can cross the laser beam or be reflected from it. We consider a planar problem when the electron bunch moves in the polarization plane of the electromagnetic wave. A general problem can be reduced to this one by an appropriate Lorentz transform \cite{rde}. In the initial frame, the problem is essentially planar so long as the momentum component normal to the polarization plane is much lesser than the component lying in it. In the case of a planar motion, the momentum components of the electrons passed the electromagnetic wave depend weakly on the initial momentum and are determined by the parameters of the laser beam and the phase of the electromagnetic wave at the entrance point of the electron. In particular, these electrons possess the identical projections of momentum to the axis normal to the laser beam. This projection is specified only by the diameter of the laser beam measured in the units of the classical electron radius. As for the reflected electrons, there is a reflection law that relates in a unique way the reflection and incidence angles. This law is independent of the parameters of the laser beam and of the ultrarelativistic electron bunch. The penetration depth of the reflected electrons to the laser beam is much smaller than the wavelength of the electromagnetic wave. Of course, these results are valid in a certain approximation that will be described in detail.

Practically speaking, we study the properties of the exact solutions to the so-called Landau-Lifshitz (LL) equation \cite{LandLifshCTF} since the exact physical solutions of the LD equation cannot be found for the field configurations we discuss. It was shown in \cite{rde} that, for such field configurations, the solutions of the LL equation provide a good approximation to the physical solutions of the LD equation even for the strong field strengths. Moreover, the solutions of the LL and LD equations tend asymptotically to each other at large times. The LL equation now becomes a standard tool in describing the evolution of electrons in a laser field with radiation reaction taken into account (see, e.g., \cite{HarGibKer,PiHaKe,HLREKR,HarHeiMar,PiMuHaKermp,SchlTikh,GBGHIKMMS}). The results of these papers also demonstrate that the behavior of electron's dynamics change qualitatively for large field strengths of the laser wave. However, these investigations are mostly numerical and cannot give a clear insight what happens in this radiation dominated regime. In many of these papers, to simplify the numerical simulations, some terms of the LL equation are thrown away (see, e.g., \cite{JPKShA,FEGNR}) without a rigorous justification, or even one dimensional simulations are only given (see, e.g., \cite{ZhKSU,NeiPiaz}). Therefore, it is of importance to analyze the exact solutions in the case when they can be found. Note in this connection that the exact solutions we shall study do not display the radiation reaction trapping effect \cite{JPKShA,FEGNR}. Though there is a considerable part of the electrons that escape the laser beam and move at small angles (less than $2^\circ$ for the intensities $I\gtrsim10^{25}$ W/cm${}^2$) to the direction of propagation of the electromagnetic wave. Another point is that certain regions (we shall accurately describe them) of the strong laser beam do become opaque for the ultrarelativistic electrons, as reported in \cite{FEGNR,EFHR}, while other regions of the beam transmit the electrons. The maximum value of the Lorentz factor for the transmitted electrons is independent of the initial momentum and is proportional to the work done by the electromagnetic fields.

\section{Notation}

We use the same notation as in \cite{LandLifshCTF}. The action functional of a charged particle with a charge $e$ and mass $m$ interacting with the electromagnetic field $A_\mu$ on the Minkowski background $\mathbb{R}^{1,3}$ with the metric $\eta_{\mu\nu}=diag(1,-1,-1,-1)$ has the form
\begin{equation}\label{action particl}
    S[x(\tau),A(x)]=-m\int{d\tau\sqrt{\dot{x}^2}}-e\int{d\tau A_\mu
    \dot{x}^\mu}-\frac1{16\pi}\int{d^4xF_{\mu\nu}F^{\mu\nu}},
\end{equation}
where $F_{\mu\nu}:=\partial_{[\mu}A_{\nu]}$ is the strength of the electromagnetic field
\begin{equation}\label{fmunu}
    F_{\mu\nu}=\left[%
\begin{array}{cccc}
  0 & E_x & E_y & E_z \\
  -E_x & 0 & -H_z & H_y \\
  -E_y & H_z & 0 & -H_x \\
  -E_z & -H_y & H_x & 0 \\
\end{array}%
\right],
\end{equation}
and we put the speed of light $c=1$.

In the natural parameterization, $\dot{x}^2=1$, the LD and LL equations read as \cite{LandLifshCTF}
\begin{equation}\label{lde_ini}
\begin{split}
    m\ddot{x}_\mu &=eF_{\mu\nu}\dot{x}^\nu+\frac23e^2(\dddot{x}_\mu+\ddot{x}^2\dot{x}_\mu),\\
    m\ddot{x}_\mu &=eF_{\mu\nu}\dot{x}^\nu+\frac23e^2(e\dot{F}_{\mu\nu}\dot{x}^\nu+e^2F_{\mu\nu}F^{\nu\rho}\dot{x}_\rho-\dot{x}^\la e^2F_{\la\nu}F^{\nu\rho}\dot{x}_\rho\dot{x}_\mu),
\end{split}
\end{equation}
respectively. The solutions of the LL equation approximate the physical solutions of the LD equation. We shall measure the lengths in the units of the Compton wavelength $l_C:=\hbar/mc$, viz., let us make a transform
\begin{equation}
    x^\mu\rightarrow l_C x^\mu,\qquad \tau\rightarrow l_C\tau.
\end{equation}
Then Eqs. \eqref{lde_ini} become
\begin{equation}\label{lde}
\begin{split}
    \ddot{x}_\mu &=\bar{F}_{\mu\nu}\dot{x}^\nu+\la(\dddot{x}_\mu+\ddot{x}^2\dot{x}_\mu),\qquad \bar{F}_{\mu\nu}:=e\hbar m^{-2}F_{\mu\nu}=\sgn(e)F_{\mu\nu}/E_0,\quad\la:=2e^2/(3\hbar)=2\al/3,\\
    \ddot{x}_\mu &=\bar{F}_{\mu\nu}\dot{x}^\nu+\la(\dot{\bar{F}}_{\mu\nu}\dot{x}^\nu+\bar{F}_{\mu\nu}\bar{F}^{\nu\rho}\dot{x}_\rho-\dot{x}^\la \bar{F}_{\la\nu}\bar{F}^{\nu\rho}\dot{x}_\rho\dot{x}_\mu),
\end{split}
\end{equation}
where $x^\mu$, $\tau$, $\bar{F}_{\mu\nu}$, and $\la$ are dimensionless quantities\footnote{Henceforth, the electromagnetic fields are the dimensionless fields $\bar{F}_{\mu\nu}$. We shall not write the bar over them.}. For the estimates, it is useful to bear in mind that all the lengths are measured in the Compton wavelength of the electron, the fields are counted in the units of the critical (Schwinger) field $E_0$, and the energies are in the units of the rest energy of the electron,
\begin{equation}
\begin{gathered}
    l_C\approx3.86\times 10^{-11}\;\text{cm},\qquad t_C\approx1.29\times 10^{-21}\;\text{s},\qquad m\approx 5.11\times10^5\;\text{eV},\\
    E_0=\frac{m^2}{|e|\hbar}\approx 4.41\times 10^{13}\;\text{G}=1.32\times 10^{16}\;\text{V/cm}.
\end{gathered}
\end{equation}
The modern accelerator facilities are able to accelerate electrons up to the energies of the order of $20$ GeV \cite{SLAC} and higher. The intensities of the laser fields, which are accessible at the present moment \cite{laser_tod,PiMuHaKermp}, are of the order of $10^{22}$ W/cm${}^2$ with the photon energies about $1$ eV. These data correspond to
\begin{equation}\label{experiment}
\begin{gathered}
    \ga\approx4\times10^4,\qquad \omega_0\approx1.47\times10^{-4},\qquad \Omega\approx1.96\times 10^{-6},\qquad\frac{\omega_0}{\Omega}\approx75.0,\\
    \la\omega_0\approx7.14\times10^{-7},\qquad\la\omega^2_0\approx1.05\times10^{-10},
\end{gathered}
\end{equation}
where $\ga$ is the Lorentz factor, $\Omega$ is the photon energy in the units of the rest energy of the electron, and $\omega_0$ characterizes the strength of the electromagnetic field in a laser beam.

\section{Dynamics of electrons in a plane wave}

Let us consider a monochromatic linearly polarized laser beam with the energy of photons $\Omega$ propagating along the $y$ axis. We can simulate such a laser beam by a plane electromagnetic wave
\begin{equation}\label{fmunu_ew}
    F^{\mu\nu}=\omega(x_-)e_-^{[\mu}e_1^{\nu]},\qquad F^2_{\mu\nu}=\omega^2e^-_\mu e^-_\nu,\qquad E_x=-\omega,\quad H_z=\omega,\qquad x\in[0,d],
\end{equation}
where $\omega(x_-)=\omega_m\cos\psi$, $\psi= \Omega x_-+\psi_0$, $x_-=x^0-y$, and $\psi_0$ is a constant phase. The four-vectors $e_-^\mu=(1,0,1,0)$ and $e_1^\mu=(0,1,0,0)$ are also introduced. Suppose that this electromagnetic field vanishes outside the strip $x\in[0,d]$, where $d$ is the diameter of the laser beam. This is a certain approximation, of course, to the real situation (for the description of real laser beams see, e.g., \cite{HSXZ}). The photons in the wave packet of the width $d$ must possess the transverse momentum of the order of $2\pi/d$. If $d\gtrsim 5\la_\ga$, where $\la_\ga=2\pi/\Omega$ is the photon wavelength, then the relative correction to the photon energy coming from the transverse momentum amounts only $k_\perp^2/(2\Omega^2)\lesssim1/50$, and so the expression \eqref{fmunu_ew} is a good approximation to the real wave packet of the width $d$. We shall assume that $d$ is of the order of $5\la_\ga$. However, as will be shown, all the results we obtain depend weakly on the beam diameter (as $d^{1/3}$).

\begin{figure}[t]
\centering
\includegraphics*[width=0.7\linewidth]{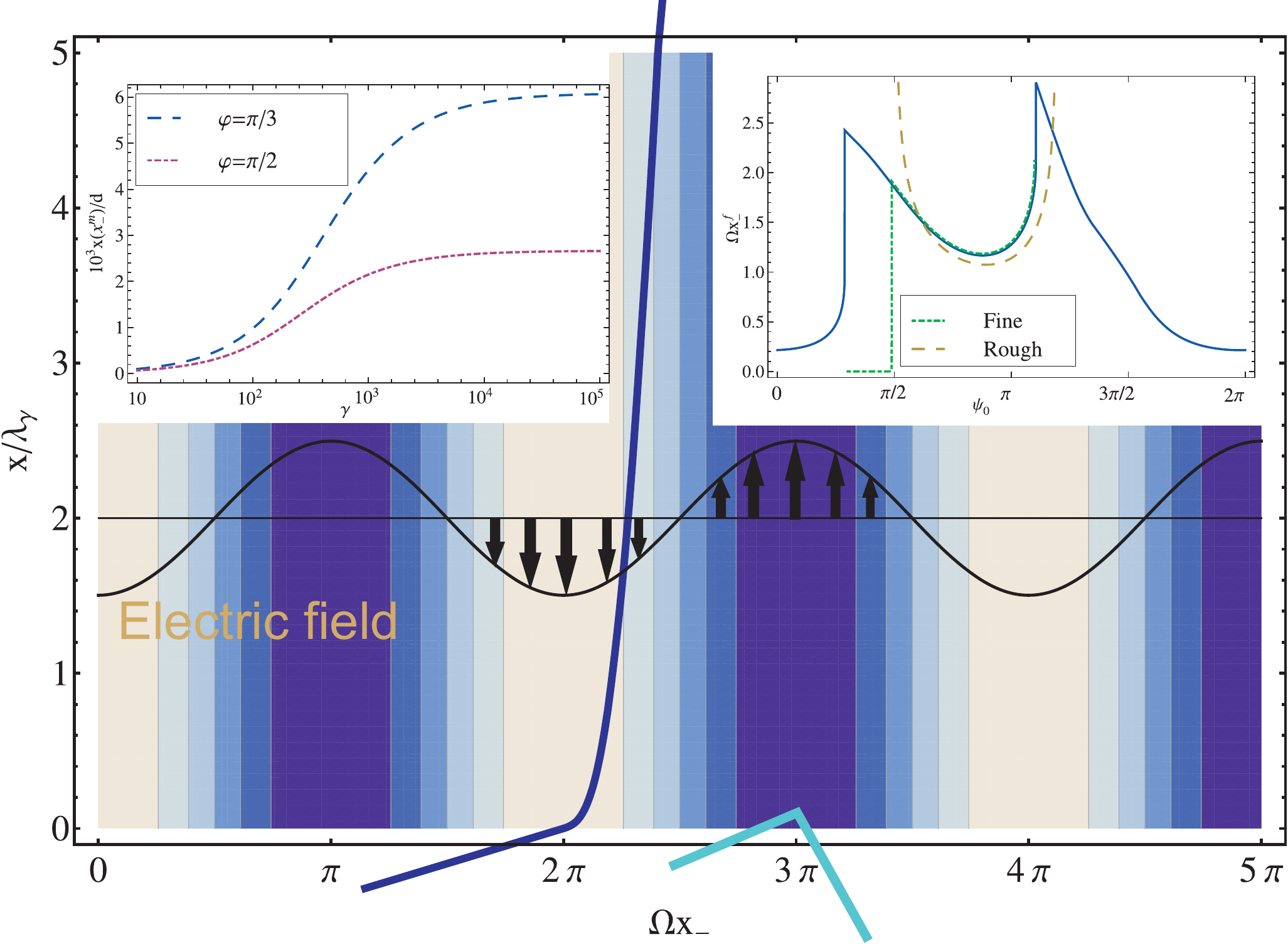}
\caption{{\footnotesize
A schematic representation of the dynamics of electrons scattering on a strong electromagnetic plane wave. The sine curve shows the electric field strength in the wave. The diameter of the laser beam is assumed to be $5\la_\ga$. The trajectory of the electron crossed the laser beam (the blue curve) is well described by formula \eqref{x_cr} without the terms depending on the initial momenta.
As for the reflected electrons (the cyan curve), the module of tangent of the exit angle is twice the tangent of the incidence angle (see the first formula in \eqref{momentum_cpts_rf}). The penetration depth is depicted schematically. The left inset: the penetration
depth in the units of the beam diameter for the different initial Lorentz factors and incidence angles at $\omega_m=10^{3/2}\omega_0$ and $\psi_0=0$. The asymptotics at large $\ga$ is in
a good agreement with \eqref{xm_fin_b}. The right inset: the value of the phase of the electromagnetic wave that the electron spends in it versus the initial phase characterizing the entrance point of the electron. The values of the parameters are $\omega_m=10^{3/2}\omega_0$, $\ga(0)=10^4$, and $\vf=\pi/2$. The rough estimate is given by \eqref{xm_fin},
while the fine estimate is obtained employing the approximate solution for $x(x_-)$, where the terms depending on the initial momenta are thrown away (the case (ii) in \eqref{cases}).}}
\label{schema_fig}
\end{figure}

As far as the electrons are concerned, we suppose they move in the plane $z=0$ in the $x$ axis direction (but not necessarily parallel to it) and strike the laser beam at the angle $\vf$ counted from the direction of propagation of the electromagnetic wave, i.e., from the $y$ axis. The exact solutions to the LD equation cannot be found in this case, but the LL equation is integrable. The solution to the LL equation has the form (see \cite{KHRL} and references there, and also \cite{NikishLLsol,Piazza,HLREKR,rde})
\begin{equation}\label{ll_sol}
\begin{split}
    \ups_-&=\Big[\ups_-^{-1}(0)+\frac{\la\omega_m^2}{4\Omega}(2\Omega x_-+\sin2\psi-\sin2\psi_0) \Big]^{-1},\\
    x_-&\Big(\ups_-^{-1}(0)-\frac{\la\omega_m^2}{4\Omega}\sin2\psi_0\Big)+\frac{\la\omega_m^2}{4\Omega^2}(\Omega^2x_-^2+\sin^2\psi-\sin^2\psi_0)=\tau,\\
    r&=r(0)-\la\omega_m\Big(1+\frac{\omega_m^2}{2\Omega^2}\Big)(\cos\psi-\cos\psi_0)-\\
    &-\Big(\frac{\omega_m}{\Omega \ups_-(0)}-\frac{\la\omega_m^3}{4\Omega^2}\sin2\psi_0\Big)(\sin\psi-\sin\psi_0)-\frac{\la\omega_m^3}{2\Omega}x_-\sin\psi+\frac{\la\omega^3_m}{6\Omega^2}(\cos^3\psi-\cos^3\psi_0).
\end{split}
\end{equation}
where $r=\ups_x/\ups_-=dx/dx_-$, $\ups_\mu=\dot{x}_\mu$ is the four-velocity, which coincides with the particle four-momentum in the system of units chosen, $\ups_-=\ups^0-\ups_y$, $\ups^0=\ga$, and we have set $x_-(0)=0$. Notice that the misprint was made in Eq. (55) of \cite{rde}: the last term in the expression for $r$ was missed. For $\la=0$ the solution \eqref{ll_sol} transforms into the well-known solution to the Lorentz equation in the plane electromagnetic wave with a linear polarization (see, e.g., \cite{LandLifshCTF}). If $\Omega x_-$ is small, the solution \eqref{ll_sol} becomes
\begin{equation}\label{ll_cr}
\begin{gathered}
    \ups_-=\Big[\ups_-^{-1}(0)+\la\bar{\omega}^2x_-(1-\Omega x_-\tg\psi_0) \Big]^{-1},\qquad x_-\ups_-^{-1}(0)+\frac{\la\bar{\omega}^2}{2}x_-^2(1-\frac{2\Omega x_-}3\tg\psi_0)=\tau,\\
    r=r(0)-\frac{\bar{\omega}x_-}{\ups_-(0)}(1-\frac{\Omega x_-}2\tg\psi_0)-\frac{\la\bar{\omega}^3}2x_-^2(1-\frac{4\Omega x_-}{3}\tg\psi_0)+\la\bar{\omega}\Omega x_-\tg\psi_0,
\end{gathered}
\end{equation}
where $\bar{\omega}=\omega_m\cos\psi_0$ and only the leading correction in $\Omega x_-$ is retained. The expression for $x(x_-)$ is obtained by integration of the expression for $r$ given in \eqref{ll_sol} or \eqref{ll_cr}. In the latter case, we have
\begin{equation}\label{x_cr}
  x=r(0)x_--\frac{\bar{\omega}x_-^2}{2\ups_-(0)}(1-\frac{\Omega x_-}3\tg\psi_0)-\frac{\la\bar{\omega}^3}6x_-^3(1-\Omega x_-\tg\psi_0)+\frac{\la\bar{\omega}}2\Omega x_-^2\tg\psi_0,
\end{equation}
where we have put $x(0)=0$. As for $r(x_-)$ presented in formula \eqref{ll_sol}, the integration over $x_-$ is readily performed, but the resulting expression is rather huge and we do not write it here. The mass-shell condition, $\ups^2=1$, implies that all the four-momentum components are expressed in terms of $\ups_-$ and $r$:
\begin{equation}\label{momentum_compts}
    \ups^0=\frac{\ups_-}2(\ups_-^{-2}+r^2+1),\qquad \ups_y=\frac{\ups_-}2(\ups_-^{-2}+r^2-1),\qquad\ups_x=\ups_- r.
\end{equation}
It is convenient to express the initial value of momentum though the Lorentz factor and the entrance angle $\vf$ of the electron to the laser beam
\begin{equation}
    \ups_-=\ga-\sqrt{\ga^2-1}\cos\vf\approx\ga(1-\cos\vf),\qquad r=\frac{\sqrt{\ga^2-1}\sin\vf}{\ga-\sqrt{\ga^2-1}\cos\vf}\approx\frac{\sin\vf}{1-\cos\vf},\qquad\vf\in[0,\pi],
\end{equation}
where it is assumed that $\ga\gg1$ and the last approximate equalities are valid for $\vf\gg\ga^{-1}$. Notice that the solutions \eqref{ll_sol} or \eqref{ll_cr} do not display the radiation reaction trapping effect \cite{JPKShA,FEGNR}. The absence of this effect for the exact solutions \eqref{ll_sol} is a consequence of the presence of a secular term (the penultimate term in the expression for $r$) \cite{KHRL,NikishLLsol}. This term dominates for $x_-$ large. The analogous term is present in the solution for a circularly polarized electromagnetic wave and also drives the electron out of the laser beam (see, e.g., Eq. (52) of \cite{rde}).

\section{Properties of the scattered electrons}

Let us turn to the properties of electrons scattered on a strong electromagnetic wave. It is clear that two cases are possible: a) the electron passes the laser beam and escapes from its opposite side, or b) the electron is reflected by the electromagnetic wave. Consider, at first, the case (a).

In order to determine the momentum components of the electron escaped, it is necessary to find the minimal positive root $x_-^f$ of the equation $x(x_-)=d$, to substitute it to the expressions \eqref{ll_sol} or \eqref{ll_cr} for $\ups_-$ and $r$, and then to employ formulas \eqref{momentum_compts}. Generally, the momentum components depend on the initial data $\ga$ and $\vf$. However, as seen from \eqref{ll_sol}, \eqref{ll_cr}, if the field strength of the electromagnetic wave is large enough, $\ga\gg1$, and $\vf\gg\ga^{-1}$, the dependence on the initial data $\ga$ and $\vf$ becomes negligibly small. The main contribution to the values of the final momentum components comes from the terms that are proportional to $\la$, i.e., from the terms that arise due to the radiation reaction effect (the so-called radiation dominated regime). The dependence on $\ups_-(0)$ and $r(0)$ is negligible provided that
\begin{equation}\label{cases}
\begin{aligned}
    i)\;\la\bar{\omega}^2x_-\ups_-(0)&\gg2&\;&\text{and}&\; 2r(0)&\ll \la\bar{\omega}^3x_-^2&\;&\text{for}&\; \Omega x_-&\ll1,\\
    ii)\;\la \omega_m^2x_-\ups_-(0)&\gg2&\;&\text{and}&\; 2\Omega r(0)&\ll \la\omega_m^3x_-&\;&\text{for}&\; \Omega x_-&\gtrsim1.
\end{aligned}
\end{equation}
Simple analytical formulas for the momenta of escaping electrons can be obtained in the case (i) only. It is this case that is realized for the strong fields and large $\ga$, since it takes a small part of the electromagnetic wave period for a charged particle to traverse the laser beam.

From \eqref{x_cr} in the case (i), we have
\begin{multline}\label{xm_fin}
    x^f_-=-\bar{\omega}^{-1}\Big(\frac{6d}{\la}\Big)^{1/3}\bigg(\frac{1-\la\bar{\omega}/(2d)\Omega x_-^f\tg\psi_0}{1-\Omega x_-^f\tg\psi_0}\bigg)^{1/3}\\
    \approx -\bar{\omega}^{-1}\Big(\frac{6d}{\la}\Big)^{1/3}\bigg(\frac{1+\la\bar{\omega}/(2d)\e}{1+\e}\bigg)^{1/3}\approx -\bar{\omega}^{-1}\Big(\frac{6d}{\la}\Big)^{1/3}(1+\e)^{-1/3}, \qquad \e=\frac{\Omega}{\bar{\omega}}\Big(\frac{6d}{\la}\Big)^{1/3}\tg\psi_0.
\end{multline}
We are interested in the case when $|\e|\lesssim1$. Then the first approximate equality in \eqref{xm_fin} is justified. The second approximate equality holds with a high accuracy for $d\gtrsim 5\la_\ga$ (see $\omega_0$ in \eqref{experiment}). In general, the last terms in \eqref{x_cr} and in the expression for $r$ in \eqref{ll_cr} can be neglected in comparison with the penultimate terms in these expressions. Henceforward, we omit these small terms. Formula \eqref{xm_fin} is valid only for the initial phases $\cos\psi_0<0$ such that the electron hits the region of the electromagnetic wave where the $E_x$ is negative (before the redefinition \eqref{lde}), i.e., the field assists the electron to cross the laser beam, see Fig. \ref{schema_fig}. Remark that $|\e|\lesssim1$ for the strengths $\omega_m\geq10^{3/2}\omega_0$ at $d$ fixed and $\Omega$ taken from \eqref{experiment}. Besides, the value of $\psi_0$ must be sufficiently far from the points $\pi/2+\pi n$, $n\in \mathbb{Z}$. Further, we shall present a more accurate estimate for the interval of phases $\psi_0$.
\begin{figure}[t]
\centering
\includegraphics*[width=0.47\linewidth]{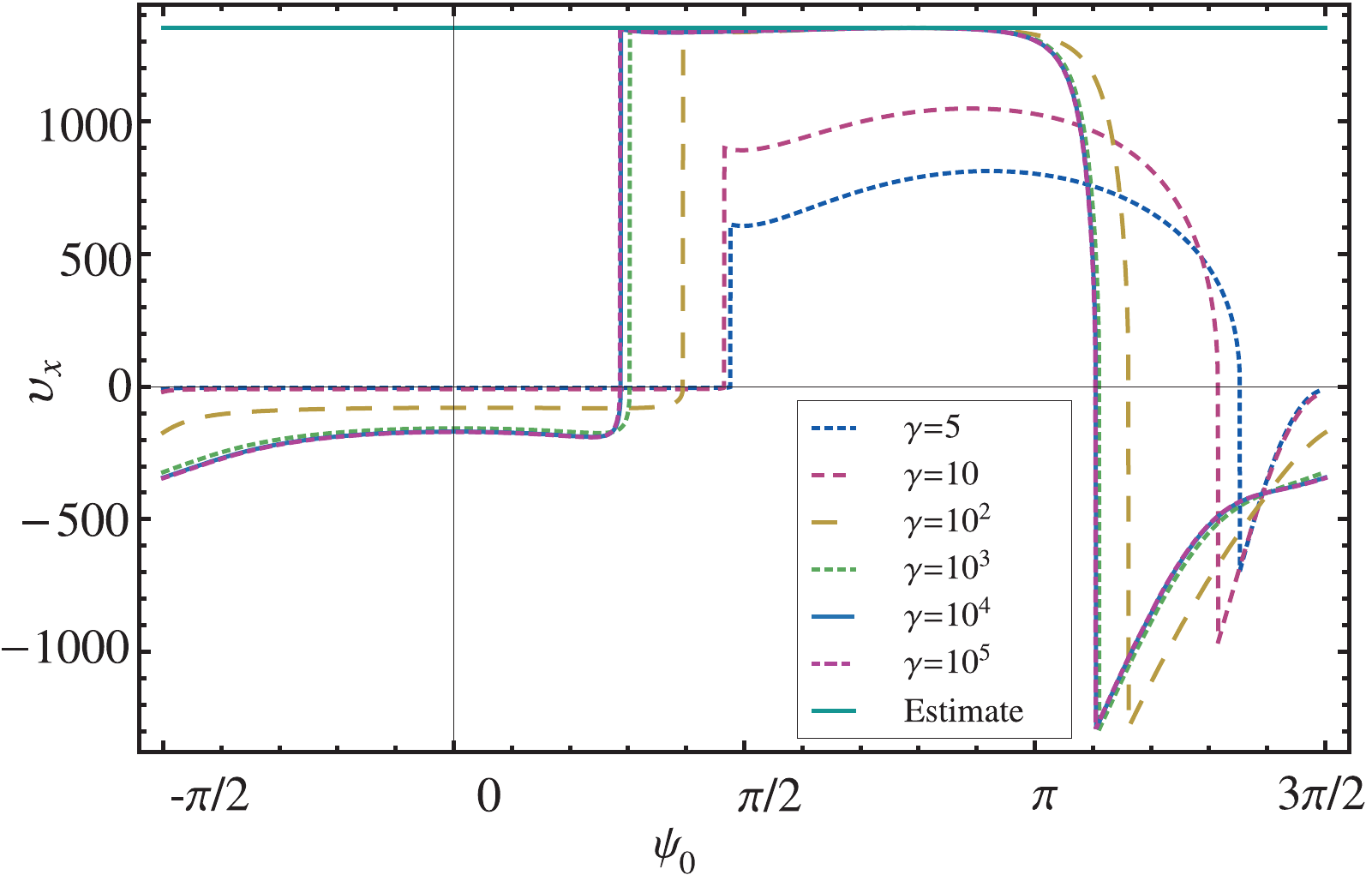}\qquad
\includegraphics*[width=0.47\linewidth]{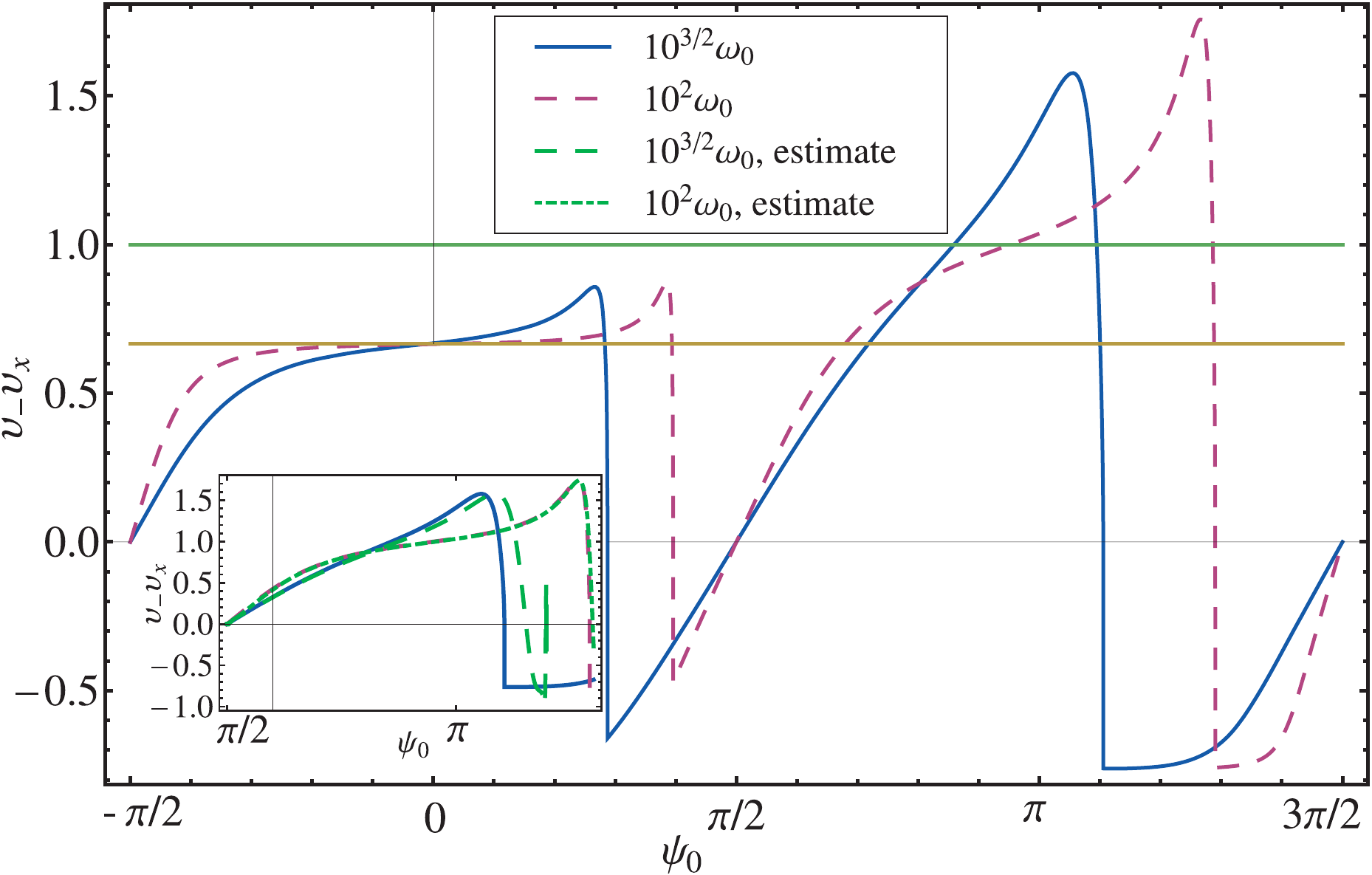}
\caption{{\footnotesize
On the left panel: the dependence of $\ups^f_x$ on the initial phase $\psi_0$ for the different Lorentz factors $\ga$ at $\omega_m=10^{3/2}\omega_0$, $d=5\la_\ga$, and the incidence angle $\vf=\pi/2$.
The straight line on the top is the rough estimate \eqref{momentum_cpts_ps} at $\e=0$. We see that this estimate holds with the a high accuracy for the electrons with $\ga(0)\gtrsim10^2$. The estimations for the zeros of $\ups_x^f$ given by formulas \eqref{phase_right}, \eqref{phase_left} are $\psi_0^r\approx1.11\pi$, $\psi_0^l\approx0.30\pi$, respectively.
On the right panel: the product $\ups_-^f\ups_x^f$ normalized on its value in a strong electromagnetic wave $(-2\la\bar{\omega})^{-1}$(see \eqref{momentum_cpts_ps}). The parameters of the laser beam are $\omega_m=10^{3/2}\omega_0$ and $\omega_m=10^{2}\omega_0$ for the solid and dashed curves on the main plot, respectively. The initial Lorentz factor $\ga(0)=10^4$ and the laser beam diameter $d=5\la_\ga$. The rough estimates obtained in \eqref{momentum_cpts_ps}, \eqref{momentum_cpts_rf} at $\psi_0=\pi$ and $\psi_0=0$ are depicted as straight lines. The latter estimation is quite good for the reflected electrons at $\omega_m=10^2\omega_0$. The larger the value of $\omega_m$ is, the flatter the plataues of the curves near these straight lines are. The inset: a comparison of the product $\ups_-^f\ups_x^f$ with the estimate (the green lines) given by \eqref{ll_sol} with $x_-^f$ taken from \eqref{xm_fin}, the terms depending on the initial momenta being cast out. The estimation is almost perfect. So the case (ii) of \eqref{cases} is realized for the transmitted ultrarelativistic electrons at these parameters of the laser beam.
}}
\label{vx_fig}
\end{figure}

Making use of \eqref{xm_fin} and the relations for the momentum components \eqref{momentum_compts}, we arrive at the approximate expressions
\begin{equation}\label{momentum_cpts_ps}
\begin{gathered}
    \ups^f_x\approx\frac12\Big(\frac{6d}{\la}\Big)^{1/3}\frac{1+4\e/3}{(1+\e)^{4/3}},\qquad \ups^f_-\approx-\frac{1}{\la\bar{\omega}}\Big(\frac{\la}{6d}\Big)^{1/3}(1+\e)^{-2/3}, \qquad\ups^f_-\ups^f_x\approx-\frac{1}{2\la\bar{\omega}}\frac{1+4\e/3}{(1+\e)^{2}},\\ \ups^f_y\approx-\frac{(1+\e)^{-2/3}}{2\la\bar{\omega}}\Big(\frac{\la}{6d}\Big)^{1/3}\Big[\la^2\bar{\omega}^2\Big(\frac{6d}{\la}\Big)^{2/3}(1+\e)^{4/3}+\frac{\la^2\bar{\omega}^2}4\Big(\frac{6d}{\la}\Big)^{4/3}\frac{(1+4\e/3)^2}{(1+\e)^{4/3}}-1\Big] \approx-\frac{3}4\bar{\omega}d\frac{(1+4\e/3)^2}{(1+\e)^2},\\
    \ups^f_0\approx -\frac{(1+\e)^{-2/3}}{2\la\bar{\omega}}\Big(\frac{\la}{6d}\Big)^{1/3}\Big[\la^2\bar{\omega}^2\Big(\frac{6d}{\la}\Big)^{2/3}(1+\e)^{4/3}+\frac{\la^2\bar{\omega}^2}4\Big(\frac{6d}{\la}\Big)^{4/3}\frac{(1+4\e/3)^2}{(1+\e)^{4/3}}+1\Big] \approx -\frac{3}4\bar{\omega}d\frac{(1+4\e/3)^2}{(1+\e)^2},\\
    \tg\al=\frac{\ups^f_x}{\ups^f_y}\approx-\frac{2}{3\bar{\omega}d}\Big(\frac{6d}{\la}\Big)^{1/3}\frac{(1+\e)^{2/3}}{1+4\e/3},
\end{gathered}
\end{equation}
where $\al$ is the exit angle of the electron, which is counted from the direction of propagation of the electromagnetic wave. The relations \eqref{momentum_cpts_ps} are valid so long as the conditions \eqref{cases} for the case (i) are fulfilled. These conditions are written as
\begin{equation}\label{val_rang}
    \la\bar{\omega}\ga\sin^2\frac{\vf}{2}\Big(\frac{6d}{\la}\Big)^{1/3}\gg1,\qquad \ctg\frac{\vf}{2}\ll\frac{\la\bar{\omega}}{4} \Big(\frac{6d}{\la}\Big)^{2/3},\qquad
    -\frac{\Omega}{\bar{\omega}}\Big(\frac{6d}{\la}\Big)^{1/3}\ll1,
\end{equation}
respectively. Numerical simulations show that the restrictions above can be weaken to a large extent, and formulas \eqref{momentum_cpts_ps} still hold with a high degree of accuracy (see Fig. \ref{vx_fig}). As follows from the general solution (see Eq. (55) of \cite{rde}) in the non-planar case, when $\ups_z\neq0$, the formulas in the first line of \eqref{momentum_cpts_ps} do not change. The other expressions in \eqref{momentum_cpts_ps} remain valid so long as
\begin{equation}
    \frac{\ups_z^2(0)}{\ups_-^2(0)}\ll r_f^2\approx \frac{\la^2\bar{\omega}^2}4\Big(\frac{6d}{\la}\Big)^{4/3}.
\end{equation}
Notice also that the mistake was made in formula (121) of \cite{rde} and in the estimations following from it since this formula does not take into account the essentially nonlinear dynamics of the electron in the asymptotic regime. In particular, the proper-time of the electron, when it escapes from the laser beam, is equal to
\begin{equation}
  \tau_{esc}\approx\frac{\la}{2}\Big(\frac{6d}{\la}\Big)^{2/3}\frac{1+2\e/3}{(1+\e)^{2/3}}.
\end{equation}
The initial data are assumed to satisfy the conditions \eqref{val_rang}. The restriction on the angles (126) of \cite{rde}, where the radiation formed on the asymptotics can be observed, becomes
\begin{equation}
  |\zeta|=\Big(\frac{k_1^2+k_3^2}{k_-^2}\Big)^{1/2}=\ctg\frac{\de}{2}\ll \frac{\la|\bar{\omega}|}{2}\Big(\frac{6d}{\la}\Big)^{2/3}\frac{1+2\e/3}{(1+\e)^{2/3}}.
\end{equation}
Here $\de$ is the exit angle of a photon counted from the $y$ axis. For $\e=0$, $d=5\la_\ga$, and $\bar{\omega}=10^{3/2}\omega_0$, the quantity standing on the right-hand side of the inequality is approximately $82.7$. In the leading order in $\e$, the spectral density of radiation in the case we consider is described by the formulas presented in section 5.1 of \cite{rde}.

\begin{figure}[t]
\centering
\includegraphics*[width=0.6\linewidth]{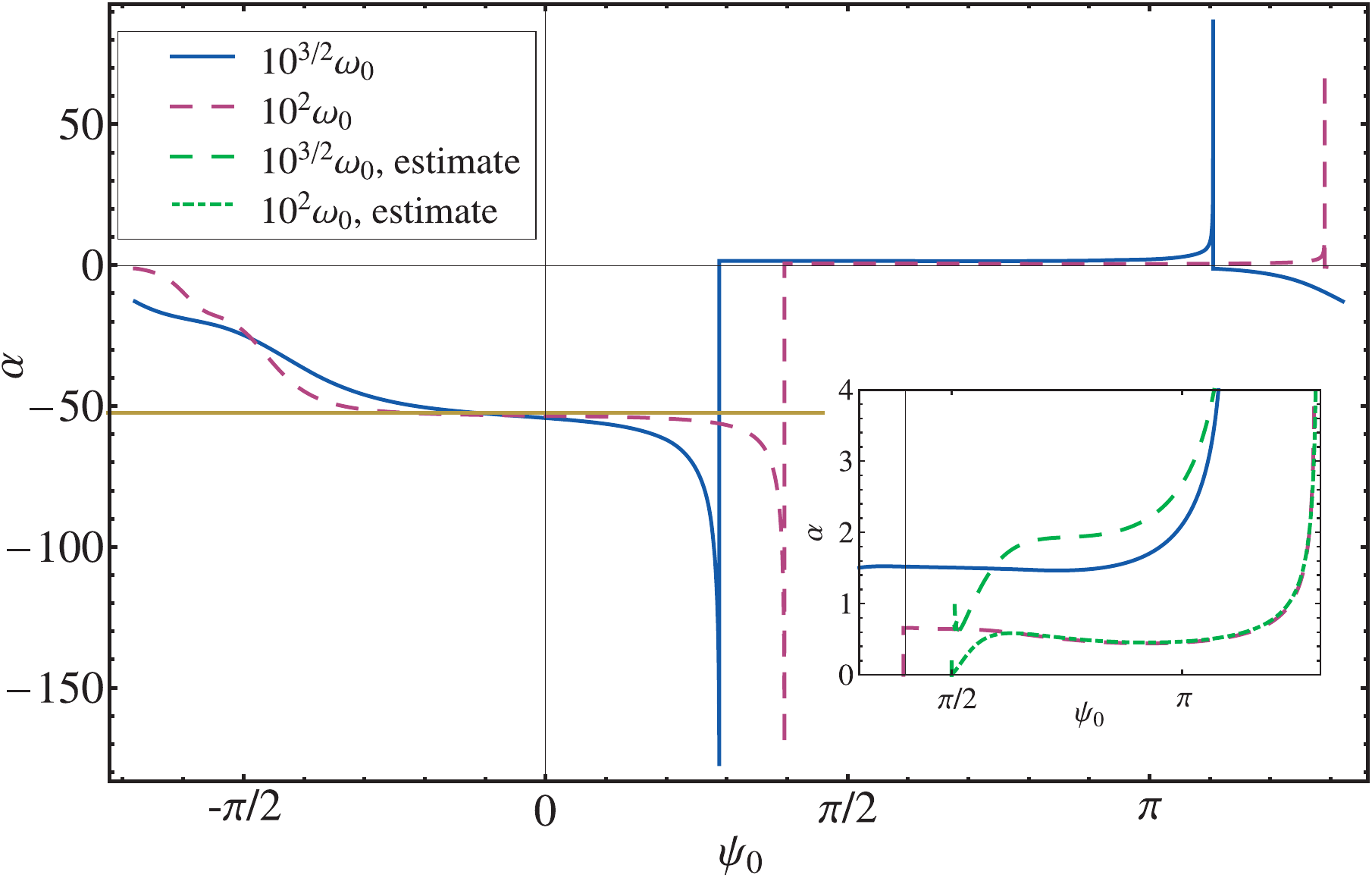}\\
\includegraphics*[width=0.48\linewidth]{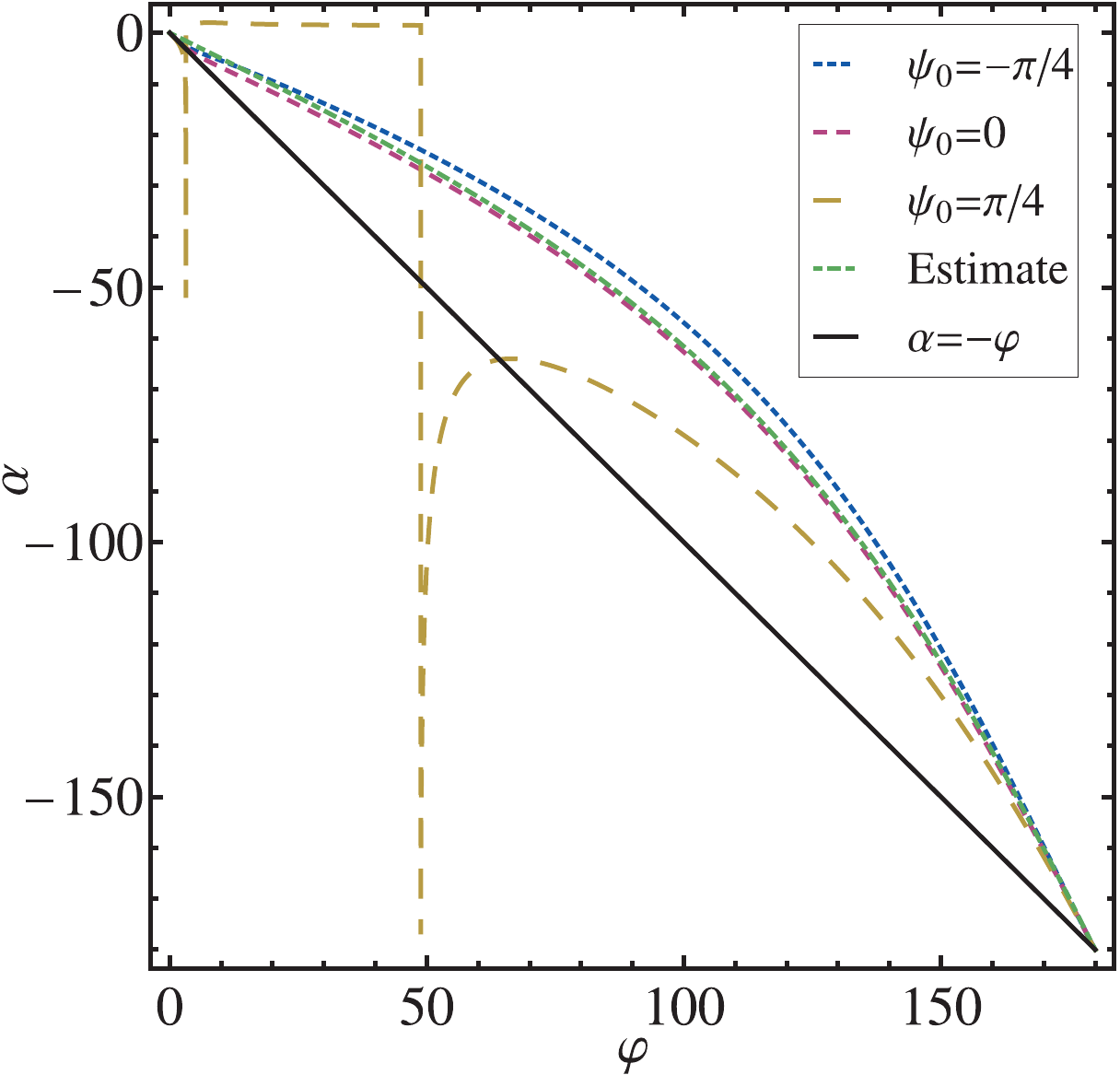}\qquad
\includegraphics*[width=0.47\linewidth]{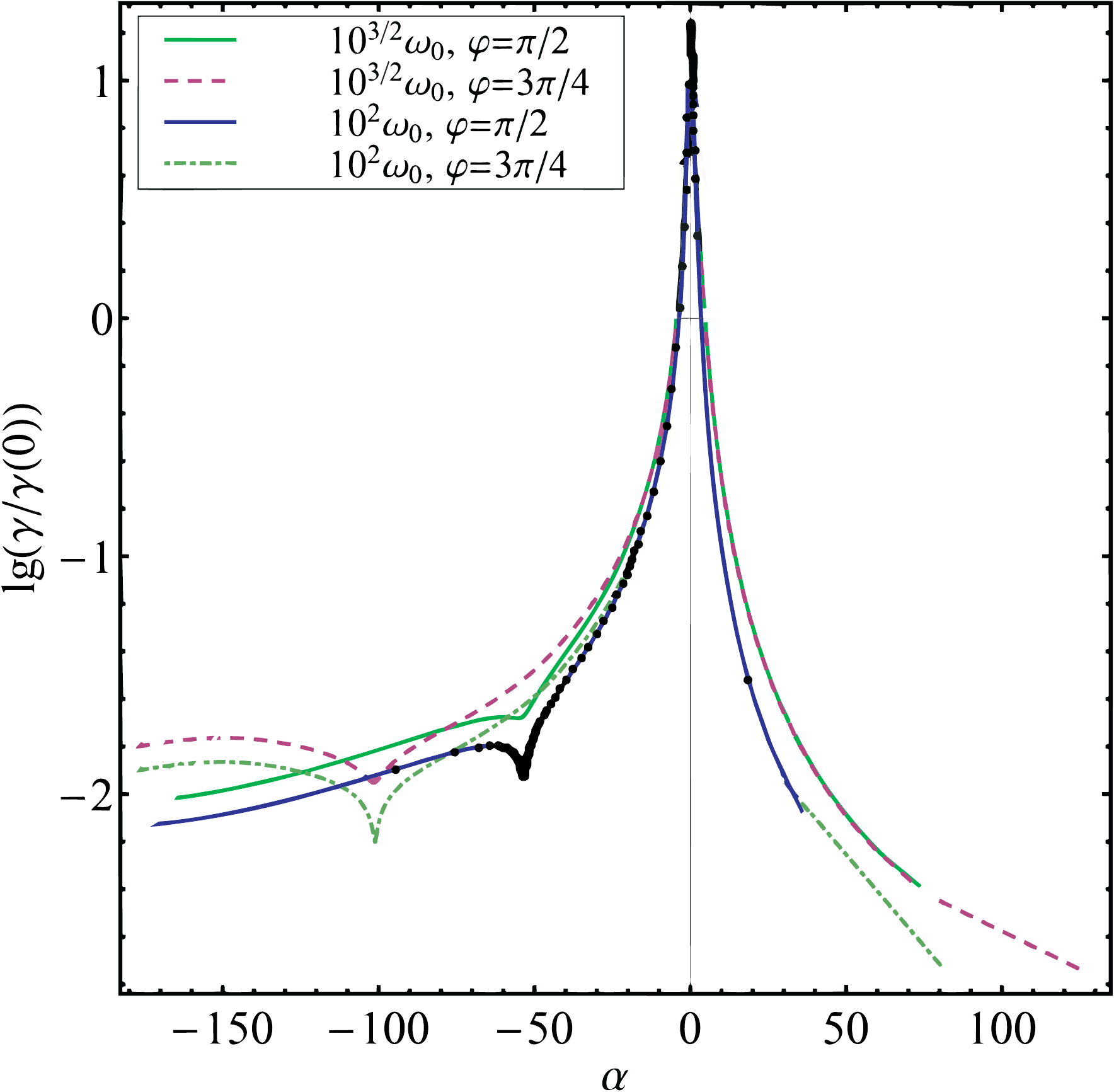}
\caption{{\footnotesize
On the top panel: the dependence of the exit angle on the initial phase $\psi_0$ for the different values of $\omega_m$. The initial data are $\ga(0)=10^4$ and $\vf=\pi/2$. The diameter of the laser beam is $5\la_\ga$. We see that the most part of the transmitted electrons escape the laser beam at small almost equal angles $\al$. The rough estimate for this angle is given in \eqref{momentum_cpts_ps} at $\psi_0=\pi$. As far as the reflected electrons are concerned, they also mostly exit at a given angle, which is determined by the reflection law \eqref{momentum_cpts_rf}. The estimate following from this law is depicted as a straight line. The inset: a comparison with the estimates (the green lines) obtained from \eqref{ll_sol} by discarding the terms depending on the initial momenta and taking $x_-^f$ from \eqref{xm_fin}. These estimates provide a rather good approximation for $\al$ at $\psi_0\gtrsim0.55\pi$. The left bottom panel:
the reflection law at $\omega_m=10^{3/2}\omega_0$ and $\ga(0)=10^4$. We see that the analytic formula \eqref{momentum_cpts_rf} for the reflection law provides a faithful estimate for the reflection angle when the phase at the entrance point $\psi_0$ lies near the plateau of the plot given on the top panel. In that case, the relative error between the exact angle deduced from the exact solution and the analytic formula \eqref{momentum_cpts_rf} is less than 2 percents. The point $\psi_0=\pi/4$ is not on this plateau and so a large deviation from formula \eqref{momentum_cpts_rf} arises.
The larger $\omega_m$ is, the lesser such deviations are. The right bottom panel: the dependence of the Lorentz factor of the escaping electrons on the exit angle for the different
strengths $\omega_m$ and the incidence angles $\vf$. The initial Lorentz factor is $10^4$ and the laser beam diameter $5\la_\ga$. The cusps on the curves at negative $\al$'s correspond to the reflection angles determined
by the law \eqref{momentum_cpts_rf}. The electrons exiting at these angles are de-excited due to the interaction with the electromagnetic wave. The cusp at $\al=0$ is formed mainly by the transmitted electrons with the Lorentz factor given approximately by \eqref{momentum_cpts_ps} at $\psi_0=\pi$. The blobs on the curve for $\omega_m=10^2\omega_0$ and $\vf=\pi/2$ are located at $\psi_0=\pi n/50$, $n\in\mathbb{Z}$. Hence, if the initial phases are equiprobable as, for example, in the case of a uniform flux of electrons impinging the laser beam, then a high density of blobs on a curve corresponds to a high probability to find electrons with such an exit angle and Lorentz factor. Therefore, the scattered electrons for the most part
either traverse the wave and exit at the small angle determined by \eqref{momentum_cpts_ps} or are reflected and exit at the angle given by the law \eqref{momentum_cpts_rf}.
}}
\label{gammaalpha_fig}
\end{figure}

We see from the relations \eqref{momentum_cpts_ps} that, up to the terms of the order of $\e^2$, the momentum component $\ups_x$ for the electrons traversed the laser beam is determined only by the diameter of the laser beam $d$ measured in the units of the classical electron radius. A numerical analysis reveals that the rough estimate ($\e=0$) for $\ups_x$ is satisfied with the accuracy of 1.4 percents for almost all the ultrarelativistic electrons crossed the laser beam (see Fig. \ref{vx_fig}). This relation takes place far beyond the bounds of validity \eqref{val_rang} of the approximations made in deriving \eqref{momentum_cpts_ps}. Also the mention should be made that, in contrast to $\ups_x$, the product $\ups^f_-\ups^f_x$ is independent of the beam diameter $d$ in the leading order in $\e$ and determined by the field strength of the electromagnetic wave at the point of entrance of the electron to the wave \cite{ldesol}. A comparison of the approximate analytic formulas \eqref{momentum_cpts_ps} with the numerical simulations is presented at Figs. \ref{vx_fig}, \ref{gammaalpha_fig}.

In the case (ii) of \eqref{cases}, the momentum components of the scattered electrons are independent of the initial values $\ga$ and $\vf$ and specified by formulas \eqref{ll_sol}, where the terms proportional to $r(0)$ or $\ups_-(0)^{-1}$ are to be omitted. The major difficulty for the analytical investigation in this case is the impossibility to solve the equation $x(x_-)=d$ in an explicit form. Though, it can be simply solved numerically. If $\Omega x^f_-\approx1$, formula \eqref{xm_fin} can be used to estimate the value of $x_-^f$. The results of numerical simulations given at Fig. \ref{schema_fig} show that the latter approximation is good enough for the field strengths  $\omega_m\geq10^{3/2}\omega_0$ at $d$ fixed and $\Omega$ taken from \eqref{experiment}.

Now we consider the case (b). In this case, the electron spends a short time in the electromagnetic wave. Hence, the ratio $r(x_-^f)$ depends on $r(0)$, while the dependence of $r(x_-^f)$ and $\ups_-(x_-^f)$ on $\ups_-(0)$ is negligibly small in the ultrarelativistic limit. The value of $\Omega x_-^f\ll1$ for almost all the reflected electrons, and so one can use formula \eqref{x_cr} to analyze the electron dynamics putting $\Omega=0$ and discarding the terms proportional to $\ups_-^{-1}(0)$ there. Then we obtain
\begin{equation}\label{xm_fin_b}
    x_-^f=\sqrt{\frac{6r(0)}{\la\bar{\omega}^3}},\qquad x_-^m=\sqrt{\frac{2r(0)}{\la\bar{\omega}^3}},\qquad x(x_-^m)=\sqrt{\frac{8r^3(0)}{9\la\bar{\omega}^3}},
\end{equation}
where $x_-^m$ is the value of $x_-$ where $x$ reaches its maximum. Obviously, formula \eqref{xm_fin_b} takes place at $\cos\psi_0>0$ only. Remark that the depth of penetration to the electromagnetic wave does not depend on the energy of the ultrarelativistic electron striking the wave, but is determined only by the field strength at the entrance point of the electron and by the incidence angle of the electron (see Fig. \ref{schema_fig}). Substituting $x_-^f$ into \eqref{momentum_compts} and casting out the terms specified above, we come to
\begin{equation}\label{momentum_cpts_rf}
\begin{gathered}
    r^f=-2r(0),\qquad\ups_-^f=(6\la\bar{\omega}r(0))^{-1/2},\qquad \ups^f_x=-\sqrt{\frac{2r(0)}{3\la\bar{\omega}}},\qquad \ups_-^f\ups^f_x=-\frac{1}{3\la\bar{\omega}},\\
    \ups_y^f=\frac{4r^2(0)+6\la\bar{\omega}r(0)-1}{\sqrt{24\la\bar{\omega}r(0)}}\approx\frac{4r^2(0)-1}{\sqrt{24\la\bar{\omega}r(0)}},\qquad \ups_0^f=\frac{4r^2(0)+6\la\bar{\omega}r(0)+1}{\sqrt{24\la\bar{\omega}r(0)}}\approx\frac{4r^2(0)+1}{\sqrt{24\la\bar{\omega}r(0)}},\\
    \tg\al=\frac{\ups^f_x}{\ups^f_y}=\frac{4r(0)}{1-6\la\bar{\omega}r(0)-4r^2(0)}\approx \frac{4r(0)}{1-4r^2(0)}\approx\frac{4\ctg(\vf/2)}{1-4\ctg^2(\vf/2)},\\
    u_x^f=\frac{\ups^f_x}{\ups^f_0}=-\frac{4r(0)}{4r^2(0)+6\la\bar{\omega}r(0)+1}\approx -\frac{4r(0)}{4r^2(0)+1}\approx-\frac{4\ctg(\vf/2)}{4\ctg^2(\vf/2)+1},
\end{gathered}
\end{equation}
where it is supposed that $r(0)\gg\la\bar{\omega}$ and $\vf\gg\ga^{-1}$ in the approximate equalities. The expressions \eqref{momentum_cpts_rf} hold provided that
\begin{equation}\label{val_rang_rf}
    \Omega\sqrt{\frac{6\ctg(\vf/2)}{\la\bar{\omega}^3}}\ll1,\qquad \sqrt{\frac{8\ctg^3(\vf/2)}{9\la\bar{\omega}^3}}<d,\qquad\ga\sqrt{6\la\bar{\omega}\sin^3(\vf/2)\cos(\vf/2)}\gg1.
\end{equation}
The second inequality (the requirement that the electron does not escape from the laser beam from its opposite side) follows from the first one at $\ctg(\vf/2)\lesssim1$ and $d\gtrsim\la_\ga$. The first restriction in \eqref{val_rang_rf} is satisfied for $\ctg(\vf/2)\lesssim1$, $\bar{\omega}\geq10^{3/2}\omega_0$, and $\Omega$ given in \eqref{experiment}. The last inequality is fulfilled at $\ga\gtrsim10^3$, $\bar{\omega}=\omega_0$, and such incidence angles that $\sin\vf\gtrsim1/10$. It is clear that the larger $\bar{\omega}_0$ and $\ga$, the larger interval of the angles $\vf$ complies with the restrictions \eqref{val_rang_rf}. The formulas in the first line of \eqref{momentum_cpts_rf} are left unchanged for a non-planar scattering, when $\ups_z\neq0$. The other relations still hold provided that
\begin{equation}
    \ups_z^2(0)/\ups_-^2(0)\ll r^2_f\approx4r^2(0).
\end{equation}

The relations \eqref{momentum_cpts_rf} for the final momentum imply that, with a high degree of accuracy, the electron is reflected from the electromagnetic wave at the angle which is determined only by the incidence angle (see Fig. \ref{gammaalpha_fig}) and is not equal to it. The electromagnetic wave transfers a momentum along the $y$ axis to the electron. As a result, the absolute value of the reflection angle is always less than the incidence angle. Inasmuch as
\begin{equation}
    \al'(\vf)=-4(3\cos\vf+5)^{-1},
\end{equation}
the field of the electromagnetic wave reflecting electrons collimates the electron bunch at the incidence angles $\vf\in[0,\vf_0]$, $\vf_0=\pi-\arccos(1/3)\approx 109.5^\circ$, i.e., having reflected, the electron bunch possesses a less dispersion of the angles. On the other hand, for $\vf>\vf_0$, the dispersion of angles in the reflected electron bunch is bigger than in the initial bunch. Notice that, analogously to the electrons traversed the laser beam, the product $\ups^f_-\ups^f_x$ for the reflected electrons is independent of the initial momentum, but equal to the different value than in \eqref{momentum_cpts_ps}. The fulfilment of this relation can be achieved easier than of the relation in \eqref{momentum_cpts_ps} (see Fig. \ref{vx_fig}). The last relation in \eqref{momentum_cpts_rf} is a consequence of the reflection law \eqref{momentum_cpts_rf} and the fact that the reflected electrons are relativistic. The approximate relations \eqref{momentum_cpts_rf} are in a good agreement with the numerical simulations (see Figs. \ref{vx_fig}, \ref{gammaalpha_fig}).

It is clearly seen at Figs. \ref{vx_fig}, \ref{gammaalpha_fig} that the values of $\psi_0$ separating the region of phases $\psi_0$, where the electrons pass through the laser beam, from the region of phases, where the electrons are reflected, are not exactly $\psi_0=\pi/2+\pi n$, as one may expect from the approximate analytic formulas \eqref{xm_fin}, \eqref{xm_fin_b}. The value of the phase $\psi_0$, where, in increasing $\psi_0$, the transmission region is superseded by the reflection region, can be roughly estimated using the expression \eqref{momentum_cpts_ps} for $\ups^f_x$. This momentum component vanishes at $\e=-4/3$. It approximately corresponds to
\begin{equation}\label{phase_right}
    \psi_0\approx\frac{3\pi}{2}-\sqrt{\frac{4\Omega}{3\omega_m}\Big(\frac{6d}{\la}\Big)^{1/3}}+2\pi n.
\end{equation}
A similar analysis for the value of the phase $\psi_0$, where the reflection is superseded by the transmission, leads to
\begin{equation}\label{phase_left}
    \psi_0\approx\frac{\pi}{2}-\Big(11.8 \frac{\Omega^2 r(0)}{\la\omega^3_m}\Big)^{1/5}+2\pi n.
\end{equation}
In order to obtain this formula, one needs to consider the dynamics of reflected particles and find such a value of the phase that $\ups_x^f=0$. Both the estimations \eqref{phase_right} and \eqref{phase_left} are in a good agreement with the numerical results.

\section{Conclusion}

Thus we have obtained several relations \eqref{momentum_cpts_ps}, \eqref{momentum_cpts_rf} among the exit momenta of the escaping electrons that are quite unexpected or may be even counterintuitive, for example, the impossibility to break through some regions of the laser beam by increasing the energy of the incident electrons. These relations were derived in making many approximation and so, at the present stage of research, can be regarded as only qualitative. It would be interesting to investigate a modification of the results above in the case of more realistic Gaussian laser beams. It is reasonable to expect that this modification will be marginal and reduced to a redefinition of the parameters $\omega_m$ and $d$. Anyway the robust results, which can be observed in an experiment, should weakly depend on a concrete form of the laser beam. Another possible direction of research is the inclusion of quantum effects such as the quantum recoil and the spin polarization of the electron wave packets. Then the results we have obtained can be used to provide a clear-cut distinction between the classical and quantum effects on the dynamics of localized electron wave packets. We leave these topics for future investigations.

\paragraph{Acknowledgments.}

We are indebt to Prof. V.~G. Bagrov for fruitful conversations. The work of OVB is supported by the RFBR grant No. 12-02-01314-a, by the Russian Federation President grant No. MK 237.2013.2, and by the grant ``Nauka'', contract No. 1.676.2014/K. The work of POK is supported in part by the Tomsk State University Competitiveness Program and by the RFBR grant No. 13-02-00551.

\end{document}